\documentstyle[12pt,aasms]{article}

\begin{document}

\def\LSUN{\rm L_{\odot}}
\def\MSUN{\rm M_{\odot}}
\def\RSUN{\rm R_{\odot}} 
\def\MSUNYR{\rm M_{\odot}\,yr^{-1}}
\def\MDOT{\dot{M}}

\newbox\grsign \setbox\grsign=\hbox{$>$} \newdimen\grdimen \grdimen=\ht\grsign
\newbox\simlessbox \newbox\simgreatbox
\setbox\simgreatbox=\hbox{\raise.5ex\hbox{$>$}\llap
     {\lower.5ex\hbox{$\sim$}}}\ht1=\grdimen\dp1=0pt
\setbox\simlessbox=\hbox{\raise.5ex\hbox{$<$}\llap
     {\lower.5ex\hbox{$\sim$}}}\ht2=\grdimen\dp2=0pt
\def\simgreat{\mathrel{\copy\simgreatbox}}
\def\simless{\mathrel{\copy\simlessbox}}

\title{On resonance line profiles predicted by radiation driven
disk wind models.}

\vspace{1.cm}
\author{ Daniel Proga$^1$} 
\vspace{.5cm}

\footnote{JILA, University of Colorado, Boulder, CO 80309-0440;
proga@colorado.edu}

\begin{abstract}
We report on resonance line profiles predicted by radiation driven disk
wind models which extend radially one order of magnitude farther out than 
our previous models. 
Our main result is that the inclusion of 
a disk wind at larger radii changes qualitatively and quantitatively
the line profiles predicted by the  models. Our new models
predict line absorption that is significantly stronger than those
predicted by old models. Some of the previous line profiles exhibit
a doubled-humped structure near the line center which is now
replaced by a more plausible single, blueshifted minimum.
We emphasize that the improvements in the shape as well as the strength
of the absorption were achieved 
without changing the gross properties of the wind. In particular,
our new models do not predict a higher mass-loss rate than the previous
models. The main changes in the line profiles
are due to the fact that the ratio between the rotational
velocity and poloidal velocity of the wind decreases downstream.
The new line profiles reproduce well the line absorption of the 
nova-like variable, IX~Vel, and promise to reproduce observations 
of other cataclysmic variables. This success of the radiation driven
disk wind model provides an important link between
outflows in OB stars and outflows in active galactic nuclei.

\end{abstract}

\keywords{ accretion, accretion disks --
           outflows  -- 
           novae, cataclysmic variables -- 
           galaxies: active --
           galaxies: nuclei -- 
           methods: numerical} 

\section{Introduction}

It appears that mass accretion onto compact objects through accretion disks 
is often accompanied by mass outflow from these disks. For example, 
disk outflows are observed in active galactic nuclei;
non-magnetic cataclysmic variables (CVs); and young stellar objects.
In the case of CVs, key evidence for outflows comes from P-Cygni profiles 
of strong UV lines such as C~IV$\lambda$1549. However, the evidence
for the outflows is not limited just to the strong UV lines (e.g.,
Long \& Knigge 2002 and references therein). 
Understanding the outflows in CVs is important because 
they have been the best observed outflows from compact objects 
and promise to provide us with insights into all disk outflows.
Until recently the interpretation of data was limited to fitting 
observed profiles to synthetic profiles calculated from kinematic
models (e.g., Mauche \& Raymond 1987; Drew 1987;
Shlosman \& Vitello 1993; Knigge, Woods \& Drew 1995; Long \& Knigge
2002). 

The most plausible mechanism for driving the CV outflows is
radiation pressure on spectral lines.
Recently, numerical hydrodynamic models of 2.5-D, time-dependent radiation 
driven disk winds have been constructed for application to
CV disk winds (Pereyra, Kallman \& Blondin 1997; 
Proga, Stone \& Drew 1998, hereafter PSD~98; 
Proga, Stone \& Drew 1999, hereafter PSD~99). 
These initial studies confirm viability of the radiation driving mechanism.
However, the same studies reveal some difficulties in obtaining
line profiles matching the observed profiles (Proga et al.
2002, Paper~1 hereafter). 
The latter is likely true even when we allow for the magnetic field
effects (Proga 2003).

We focus here on re-examining pure radiation-driven disk wind models
for CVs. In particular, we examine the size of the region where
the line is formed and show that what is required to improve
the model predictions is to capture the entire line formation region
by increasing the radial range of the models. 
In Section~2, we summarize the key elements of our calculations.
The results are presented in Section~3. 
In Section~4, we discuss key aspects of the synthetic line profiles and make
comparisons with recent observations.  

\section{Method}

For this paper, we recalculated dynamical models presented by PSD~99
and calculated a few new models.
We refer a reader to PSD~99 for details
on wind models and to Paper~1 for details on line-profile calculations.
The computations of line profiles are exactly as in Paper~1.
We made only two minor modifications to the models:
(i) we increased the outer radius of the computational domain, $r_o$,
from 10 to 100 white dwarf radii and (ii) as in PSD~99, we 
computed models with the radiation force
using the intensity of the radiation
integrated over all wavelengths, $\lambda$, 
but also using the intensity of the radiation
integrated over the UV band only (i.e., $200~\AA \leq \lambda \leq
3200~\AA$).
We call the former $I$ and the latter $I_{\rm UV}$.
We found that the second modification can change
the wind solution (e.g., there is no outflow originating from
large radii on the disk)
but typically this is not much reflected in the predicted line profiles.
On the other hand, the first modification
does not change the wind solution but has a significant
effect on the predicted line profiles. 

\section{Results of Profile Calculations}

Our basic set of line profiles is based on four hydrodynamical models computed
on the larger computational domain with the model parameters as in 
PSD~99's  
             models A, B, C, and D (see Table 1).
We call our models 
recalculated with the intensity $I$ on the larger
grid  A1, B1, C1, D1, respectively, while models
recalculated with the intensity $I_{\rm UV}$ 
on the larger grid are labeled A2, B2, C2, D2.
To examine the effect of viewing angle on the line profiles,
for each disk wind model we compute line profiles for five
inclination angles: $i=~5^\circ, 30^\circ, 55^\circ, 70^\circ$, and 
$85^\circ$. Examining effects of inclination angle is important because of the 
strong inclination angle dependence observed in CV spectra.
Figure~1 shows some of the  model line 
profiles obtained after scaling to unit continuum level.  

To illustrate the key differences between profiles from Paper~1
and our profiles, we present results from the B models and the C models.
We mention however,
that the changes in the line profiles for the A models and the D models 
are qualitatively similar to the changes for the B models and 
the C models.

Table~1 summarizes the main input parameters of the hydrodynamical disk
wind models including the mass accretion rate, $\MDOT_a$, total stellar
luminosity relative to the disk luminosity, $x$, and resultant total system
luminosity, $L_{tot}=L_D+L_\ast=(1+x) L_D$. Additionally, the table
lists, the gross 
properties of the disk winds including the mass-loss rate, $\MDOT_w$, 
characteristic velocity of the fast stream at 10$r_\ast$, $v_r(10r_\ast)$ 
for model A, B, C, and D and their counterparts for the models based on the
larger grid,  $v_r(100r_\ast)$, and flow opening angle, $\omega$.  

First, we present our results for a wind for which
$\MDOT_a=\pi\times10^{-8}~\MSUNYR$ and the white dwarf (WD) is assumed to 
be dark $x=0$ (model~B, B1, and B2). 
We note that the B  models have similar properties,
except for $\MDOT_w$ which is lower for B2 than for models B and B1
(see Table~1).
The lower $\MDOT_w$ in model B2 is due to the fact
that the radiation force in model B2 is lower than for models B and B1
as it is computed using
$I_{\rm UV}$, which is lower than $I$. 
The absence of a wind model A2 is caused by the same
effect, at a more extreme level. The A models are for a relatively low 
luminosity and a reduction of the driving radiation flux resulting 
from taking into account only the UV photons makes the total
radiation force too weak to overcome the gravity and drive a wind.

The five top panels of Figure~1 (Figs. 1a-1e) compare the profiles
for  model~B as a function of $i$ from Paper~1
(thin solid lines) with the profiles predicted by models B1 and B2
(dashed and thick solid lines, respectively).
The line profiles from Paper~1 for $i=55^\circ$ and $70^\circ$ 
show two maximum absorptions almost equally shifted from the line center 
to the blue and red. 
This effect is due to the fact that
the rotation dominates over expansion in shaping the line profile for 
$r\simless 10 r_\ast$. There is little absorption near the line center. 
The most obvious change in the profiles 
caused by taking into account the wind at larger radii
is the increase of the absorption near the line center.
Our detailed analysis of the line-forming region shows
that this central absorption is due to the wind originating
at $3 \simless r \simless 10~r_\ast$ which has optically thick resonance 
surfaces at $r\simless 30~r_\ast$  to the photons close to the line center.
Thus, the `third' component of the absorption due to the 
wind at large radii changes
qualitatively the shape of the total line profile for $i=55^\circ$ 
and $70^\circ$ and resolves the problem of the two maximum
absorptions.
For $i \simless 30^\circ$, the outer wind 
significantly contributes the line absorption 
at all wavelengths.

We note that the red and blue edges of the lines
do not change when the  wind at larger radii is taken into account
(compare thick solid lines with dashed and thin solid lines in Fig.~1).
This reflects the fact that the old and new wind solutions are very similar 
within $10~r_\ast$ and that the old models
capture well the acceleration zone of the fastest part of
the wind (e.g., PSD~98). Note that the fastest wind is 
responsible for the position of the blue edge of the absorption
at the lowest inclination.

The line profiles based on our new models have a similar
dependence on the inclination angle as the line profiles from 
Paper~1. Namely, 
the absorption weakens whereas
the scattered emission strengthens as inclination angle increases.
Additionally, the red edge is at the line center
for $i=5^\circ$ and moves to the red as $i$ increases for $i<85^\circ$.

The five bottom panels of Figure~1 (Figs. 1f-1j) compare the profiles
for model~C as a function of $i$ 
(thin solid lines) with the profiles predicted by models C1  and C2
(dashed and thick solid lines, respectively).
Model~C is an example of a strong steady state wind where  
$\MDOT_a=\pi\times10^{-8}~\MSUNYR$ and WD radiates at the same
rate as the disk, $x=1$. 
In analogy with the B models, the C  models have similar properties
except for $\MDOT_w$, which is lower for C2 than for model C and C1
(see Table~1).

The line profiles for model~C (fig.1f-1j) are different from those for 
model~B for $i> 30^\circ$. 
For model~B, the line  develops a strong, more or less symmetric, 
double-peaked emission when seen nearly edge-on whereas for model~C, the
line  develops a typical P~Cygni profile with blueshifted absorption and 
redshifted emission. Additionally, the line is significantly broader 
for model~C than for model~B.
The rotation of the wind does not dominate over expansion in shaping the line.
In particular, the shape and strength of the blueshifted absorption is due
to expansion. For $i\simless 55^\circ$, there is 
a rotationally broadened redshifted absorption.
The redshifted absorption is weaker and narrower 
than the expansion-dominated blueshifted absorption. 

The contribution from the outer wind causes similar
changes to the line profiles for model~C as it does to the profiles
for model~B.
The line profiles for model~C from Paper~1 show
a deep narrow absorption at $v\sim - 1000~{\rm km~s^{-1}}$ for $i=70^\circ$ 
(Fig. 1i) and at $v\sim - 200~{\rm km~s^{-1}}$ for $i=85^0$ (Fig. 1j).
The central absorption due to the outer wind broadens and strengthens
the absorption for $i=70^\circ$ and eliminates the central emission
which has a peak at $v\sim - 500~{\rm km~s^{-1}}$ in the line profile
from Paper~1. Additionally, the outer wind
strengthens the red-shifted emission for $v\simless 1500~{\rm km~s^{-1}}$.
As a result, for $i=70^\circ$, the line profiles predicted by
models C1 and C2 does not exhibit 
the double-humped structure seen in the line profile predicted by
model~C.
For $i=85^\circ$, the line profiles predicted by
models C1 and C2 develop a double-peaked emission without 
the deep narrow absorption at $v\sim - 1000~{\rm km~s^{-1}}$ 
which is present in the profile from Paper~1. This narrow
absorption disappeared in the profiles for model C1 and C2,
mainly because the blue-shifted absorption
is weaker in our new model than in the old model.

For $i<70^\circ$, the line profiles for  model~B and C 
are similar in three respects: (i) the scattered emission is weak,
(ii) the contribution from the outer wind
is negligible in the red part  but it is significant
in the blue part of the profiles, and (iii) 
the contribution from the outer wind does not change the position of 
the red and blue edges of the lines.

In analogy to models B and C, at $i=85^\circ$, 
the scattered emission is seen as weaker 
by the observer for  models C1 and C2  than for models B1 and B2 
because the background continuum
emission is stronger in the C models. 
At this high inclination, the background continuum of the C model
is dominated by the WD contribution
(see Table~1 in Paper~1). 

\section{Discussion and comparison with observations}

Our main result is that the inclusion of  a
disk wind at larger radii changes qualitatively and quantitatively
the line profiles predicted by the pure radiation-driven disk wind model. 
The models computed on a small grid -- such as  those in PSD~99, 
where the outer radius equals 10 $r_\ast$ --
are sufficient to calculate the gross properties of the disk wind. 
For example, PSD~99's simulations are sufficient to calculate
the wind mass-loss rate and the fast part of the  wind, which 
are both associated with  the outflow from 
the innermost disk.
However, the previous simulations
do not capture the entire region where lines are formed.
As a result, they underpredict the line absorption and to a lesser
extent the scattered emission.
The previous simulations predicted a double-humped structure near the line
center for intermediate inclinations. 
This structure is due to a nonnegligible red-shifted absorption
that is formed in the slow  wind where the rotational
velocity dominates over expansion velocity. We showed here that by taking
into account the downstream part of the same slow  wind 
we were able to increase significantly the central absorption.
As a result, the double-humped structure is reshaped to a more
typical broad trough. We emphasize that all improvements
in the shape as well as the strength of the absorption were 
achieved without changing the gross properties of the wind. In particular,
our new models do not predict a higher mass-loss rate than the previous
models. The changes in the line profiles
are mainly caused by the fact that the ratio between the rotational
and poloidal velocity decreases downstream.

In Paper~1, we showed that the double minimum apparent near line center can 
be replaced by a more plausible single, blueshifted minimum when we make 
the slow wind transparent. We concluded then, that the wind-formed line 
profiles seen at ultraviolet wavelengths cannot originate in a flow where 
rotation and poloidal expansion are comparable. 
The UV lines must trace gas that expands substantially faster 
than it rotates. Our new results support this reasoning.

Our new and old line profiles have many features that are qualitatively 
consistent with the features of the observed line profiles.
These include the classical P-Cygni shape for a range of 
inclinations, the location of the maximum depth of the absorption 
component at velocities less than the terminal velocity (a requirement 
discussed in Drew 1987), and the transition from net absorption to net 
emission with increasing inclination.  However, our new line profiles
are also qualitatively consistent with the observations,
in particular, the depth and width of the absorption.
The main discrepancy between the predicted line profiles
and the observed ones is in the line emission. Specifically, the model cannot
produce the redshifted emission as strong as that seen, for example,
in the C{\sc iv} profile of many systems 
with intermediate inclinations (see below). However, this shortcoming  
is not a great surprise -- this has been a problem for a while 
(see e.g. Mauche, Lee \& Kallman 1997; Ko et al. 1998).

We are undertaking a systematic comparison between predicted line profiles and 
observations for many systems. Such  a study is needed to support
our finding that the pure radiation-driven disk wind model 
can work for CV winds in the sense that it can reproduce the observed
line profiles for model parameters (e.g., system luminosities)
suitable to CVs. Our preliminary results from 
limited survey of dynamical models and their predictions
are promising.
To illustrate this, we compare our new profiles with 
spectrally well-resolved UV observations for the brightest 
nova-like variable, IX~Vel (Hartley [2002]; in Paper~1, 
we made a model comparison with the same data).
In Figure 2, we show a comparison between profiles derived
from the model with $\MDOT_a=\pi\times10^{-8}~\MSUNYR$, $x=0.25$ (model E2) and
observations of the C{\sc iv}~1549\AA\ and Si{\sc iv} 1403\AA\ transitions.
To show how much line emission is required we 
plot only the absorption component of the line 
synthesized for $i =50^\circ$ and $60^\circ$.
The observed $i$  for this system is  
$60^\circ$   (Beuermann \& Thomas 1990).
The choice for the relative luminosity of WD (i.e., $x=0.25$) 
was motivated by the fact that
the line profiles predicted by model B2 with $x=0$ and model C2 
with $x=1$  bracket the observed profiles.

Figure 2 clearly shows that the model profiles well
reproduce  the blue-shifted absorption. Therefore, 
we conclude that we have significantly narrowed
the gap between the kinematics of the radiation-driven models 
and reality. The gap is only narrowed but not bridged yet
because the mass fluxes required to match the observed spectra at least
of IX~Vel are somewhat higher than those observed. In particular,
it is believed that the luminosity of the system requires
a mass accretion rate of at most $10^{-8}~\MSUNYR$, whereas
our line profiles require $\MDOT_a=\pi\times10^{-8}~\MSUNYR$ and $x=0.25$,
yielding a system luminosity higher than the observed one by 
a factor of $\sim 4$. However, our main point is that
this discrepancy is much smaller than it used to be 
(i.e., it was more than 1 order of 
magnitude) and we believe that it can be reduced still further.
We have computed many models, changing various model
parameters such as $\MDOT_a$ and the parameters of the force multiplier,
$\alpha$ and $M_{max}$. 
In general, we find that there is a degeneracy
in the model parameters as far as line profiles are concerned
(PSD~98 found an analogous degeneracy for the wind properties).
For example, models with slightly different parameters
-- such as $\MDOT_a$ and $\alpha$ -- produce similar line profiles 
for different $i$. Additionally, 
a model with the same parameters as model A2 but with $\alpha=0.674$
instead $\alpha=0.6$ predicts very similar line profiles
to model B2 for the same $i$. Finally, we find that
the product $(1+x) L_D M_{max}$, 
not  its individual factors, appears to be a fundamental parameter
determining the line profiles (i.e., their width and depth)
for the parameter range applicable to CVs. This has an important
implication for our models: to obtain a theoretical fit as good
as shown in Fig. 2 for a fixed $i$, 
we need $(1+x) L_D M_{max} \sim 1.3 \times 10^5~\LSUN$
rather than specifically $x=0.25$, $M_{max}=4400$, and $L_D=23.4~\LSUN$
as for model E2.

As in Paper~1, some aspects of our results depend on our
assumptions about the disk and microphysics of the wind.
Further work should include in particular, self-consistent calculations 
of the wind ionization state.

ACKNOWLEDGMENTS: 
We thank J.E. Drew and M.C. Begelman for their comments.
We acknowledge support from NASA under LTSA grants NAG5-11736 and NAG5-12867.
We also acknowledge support provided by NASA through grant  AR-09532
from the Space Telescope Science Institute, which is operated 
by the Association of Universities for Research in Astronomy, Inc., 
under NASA contract NAS5-26555.

\newpage
\section*{ REFERENCES}
 \everypar=
   {\hangafter=1 \hangindent=.5in}

{

  Beuermann, K., Thomas, H.-C., 1990, A\&A, 230, 326

  Drew, J.E. 1987, MNRAS, 224, 595

  Hartley, L. E., Drew, J. E., Long, K. S., Knigge, C., \& Proga, D. 2002, 
  MNRAS, 332, 127 

  Knigge, C., Woods, J.A., Drew, J.E. 1995, MNRAS, 273, 225

  Ko, Y., Lee, P. Y., Schlegel, E. M., \& Kallman, T. R. 1996, ApJ, 457, 363 

  Long, K.S., \& Knigge, C., 2002, ApJ, 579, 725

  Mauche, C.W., Lee, Y.P., Kallman, T.R. 1997, ApJ, 477, 832

  Mauche C.W., Raymond J.C. 1987, ApJ, 323, 690

  Mauche C.W., \& Raymond J.C. 2000, ApJ, 323, 690

  Pereyra, N.A., Kallman, T.R., \& Blondin, J.M. 1997, ApJ, 477, 368

  Proga, D. 2003, ApJ, 585, 406

  Proga, D., Kallman, T.R., Drew, J.E., \& Hartley, L.E., 2002, ApJ, 572,
  382 (Paper 1) 

  Proga, D., Stone J.M., \& Drew J.E. 1998, MNRAS, 295, 595 (PSD~98)

  Proga, D., Stone J.M., \& Drew J.E. 1999, MNRAS, 310, 476 (PSD~99)

  Shlosman I., \& Vitello P.A.J. 1993, ApJ, 409, 372

}

\eject

\newpage
\centerline{\bf Figure Captions}

\vskip 4ex
\noindent
Figure~1 -- Line profiles for hydrodynamical disk wind models as a function
of inclination angle, $i$ (see top right corner of each panel 
for the value of $i$).
The top panels, a-e, are for models with $x=0$ but with $\MDOT_{a} = 
\pi \times 10^{-8}~\MSUNYR$  (B models).
The bottom panels, f-j, show results for models with 
$\MDOT_{a} = \pi \times 10^{-8}~\MSUNYR$ and
$x=1$ (C models).
The thin solid lines are the line profiles based on the 
models B and C from PSD~99 (these are the same profiles as in Paper~1,
see figure 1 there). The dashed lines 
are line profiles based on our models B1 and C1 
while the thick solid lines are profiles based on models B2 and C2.
The zero velocity corresponding to the line center is indicated by the vertical
line.
Note the difference in the velocity and flux ranges in the planes
for $i=85^\circ$ (fifth column).

\vskip 4ex
\noindent
Figure~2 -- 
Comparison between profiles derived
from model E2 for $i=50^\circ$ and $60^\circ$
(thick solid and thick dashed line, respectively) and 
observations of the C{\sc iv}~1549\AA\ and Si{\sc iv} 1403\AA\ transitions in 
the spectrum of the brightest nova-like variable, IX~Vel 
(Hartley et al. [2002]).  The synthesized lines 
show the absorption component without the contribution from
the scattered emission (see the main text).

\newpage

\begin{figure}
\begin{picture}(180,590)

\put(0,0){\includegraphics{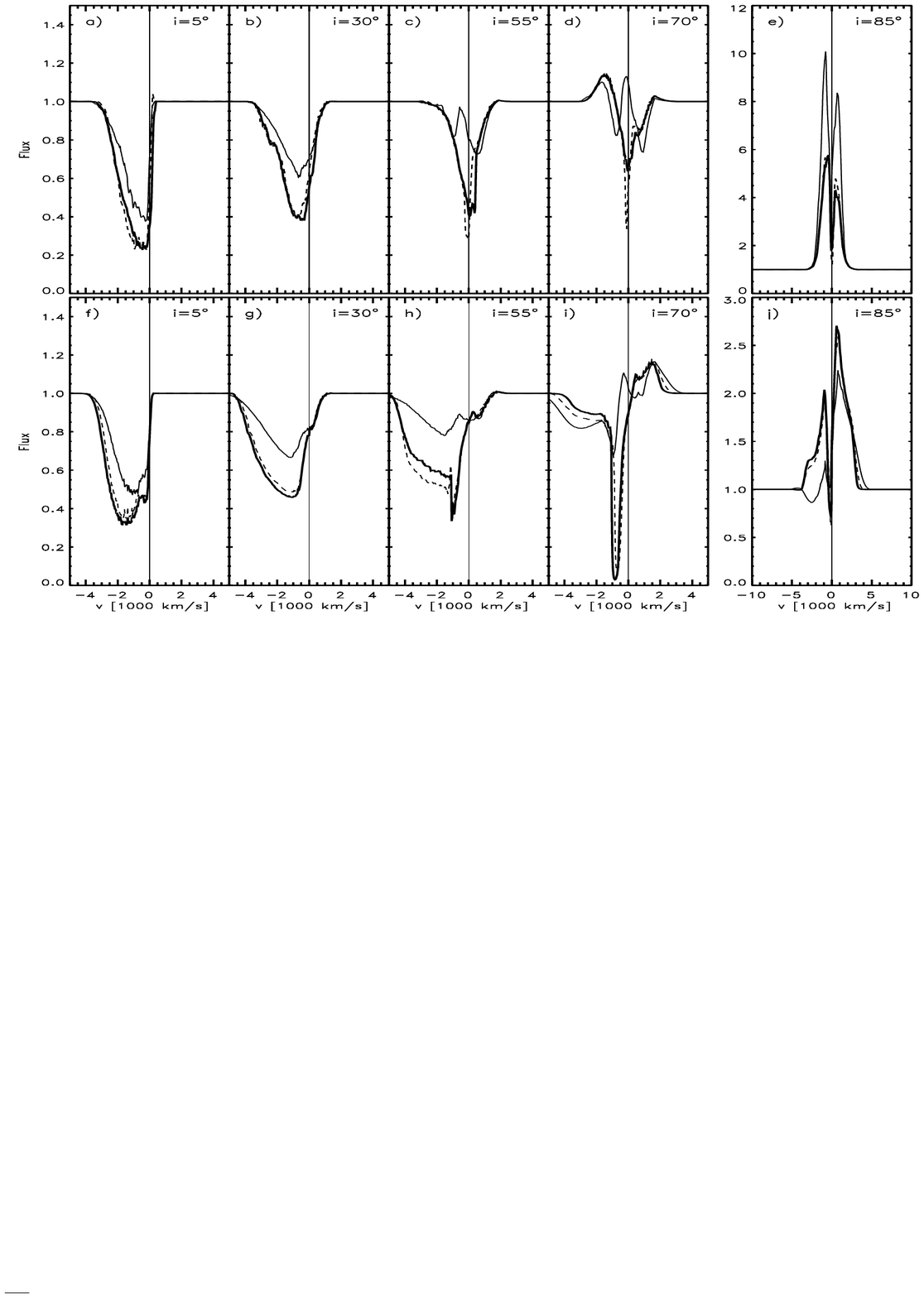}}

\end{picture}
\caption{}
\end{figure}

\begin{figure}
\begin{picture}(180,590)

\put(280,423){\includegraphics{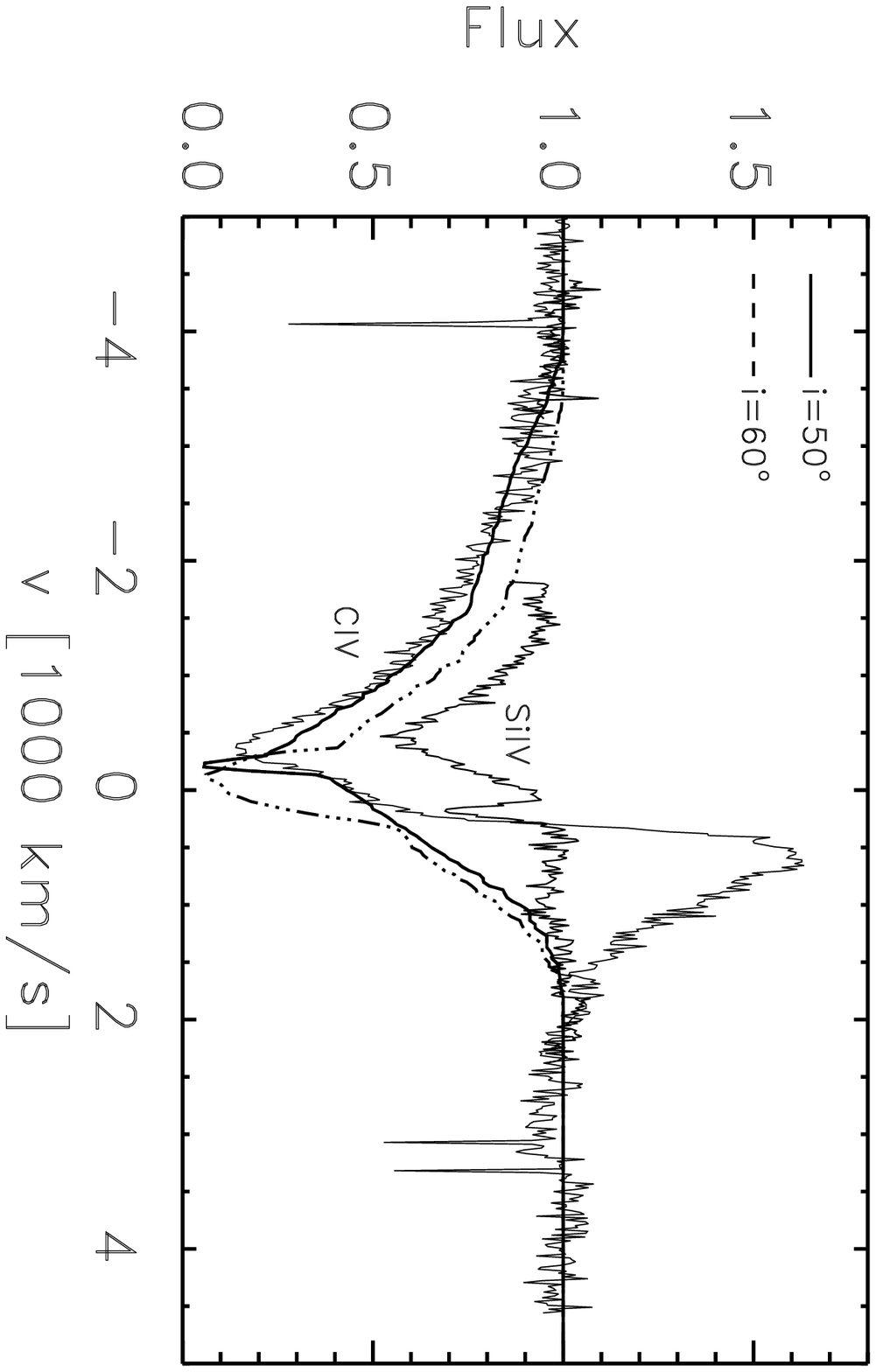}}

\end{picture}
\caption{}
\end{figure}

\eject

\begin{table*}
\footnotesize
\begin{center}
\caption{Input parameters and main characteristics of the hydrodynamical models used in this paper
(for models A, B, C, and D see also PSD~99)}
\begin{tabular}{c c c r c r r c } \\ \hline
     &                         &   &       &                          & &            &      \\
 model & $\MDOT_a$          & $x$ & $L_{tot}$ & $\MDOT_w$  &   $v_r(10 r_\ast)$ & $v_r(100 r_\ast)$ & $\omega$   \\
     & (M$_{\odot}$ yr$^{-1}$) &   & $\LSUN$  & (M$_{\odot}$ yr$^{-1}$)  & $(\rm km~s^{-1})$ &  $(\rm km~s^{-1})$ & degrees \\ \hline

     &                         &   &       &                          & &            &      \\
A    &$  10^{-8}$              & 0 & 7.5   &  $ 5.5\times10^{-14}$    & 900  & --    &  50  \\
B    &$ \pi \times 10^{-8}$    & 0 & 23.4  &  $ 4.0\times10^{-12}$    & 3500 & --    &  60  \\
C    &$ \pi \times 10^{-8}$    & 1 & 46.9  &  $ 2.1\times10^{-11}$    & 3500 & --    &  32  \\
D    &$ \pi \times 10^{-8}$    & 3 & 93.8  &  $ 7.1\times10^{-11}$    & 5000 & --    &  16  \\
     &                         &   &       &                          & &            &      \\
A1    &$  10^{-8}$              & 0 & 7.5   &  $ 5.5\times10^{-14}$    & 900  & 900      &  55  \\
B1    &$ \pi \times 10^{-8}$    & 0 & 23.4  &  $ 4.0\times10^{-12}$    & 3500 & 3500    &  65  \\
C1    &$ \pi \times 10^{-8}$    & 1 & 46.9  &  $ 2.1\times10^{-11}$    & 3500 & 3500    &  45  \\
D1    &$ \pi \times 10^{-8}$    & 3 & 93.8  &  $ 7.1\times10^{-11}$    & 5000 & 5000    &  22  \\
     &                         &   &       &                          & &            &      \\
A2    &$  10^{-8}$              & 0 & 7.5   &       --                 & 800  & --  & --      \\
B2    &$ \pi \times 10^{-8}$    & 0 & 23.4  &  $ 2.2\times10^{-12}$    & 3500 & 3500    &  65  \\
C2    &$ \pi \times 10^{-8}$    & 1 & 46.9  &  $ 1.9\times10^{-11}$    & 3500 & 3500    &  45  \\
D2    &$ \pi \times 10^{-8}$    & 3 & 93.8  &  $ 6.3\times10^{-11}$    &
5000 & 5000    &  30  \\
     &                         &   &       &                          & &            &      \\
E2    &$ \pi \times 10^{-8}$    & 0.25 & 29.3  &  $ 8\times10^{-12}$    & 3500 & 3500    &  55  \\
\hline

\end{tabular}
\end{center}
\end{table*}
\eject

\end{document}